\documentstyle[aps,preprint]{revtex} \baselineskip 12pt
\begin{document}
%\wideabs{
%\draft
\title{ Can gluon condensate in pulsar cores  explain pulsar glitches ?}

\vspace{2.5 in}

\author{ Raka D. Ray
%\footnote{email : raka@rri.ernet.in}
}
\address{ Raman Research Institute,  Bangalore 560 080,  India \\e-mail address:- raka@rri.ernet.in}

\vspace{.2in}

%\twocolumn

\maketitle

\vspace{.2in}

\begin{abstract}

Making use of 
 the possibility that gluon condensate  can be formed in
neutron star core, we study 
the vortex pinning force
between the crust and the interior of the neutron star.
 Our estimations indicate an increase in pinning strength
with the age of the neutron star. This helps in explaining
observed pulsar glitches and removes some difficulties
faced by vortex creep model.

\end{abstract}

\newpage

Pulsars are considered as strongly magnetized rotating neutron stars 
 emitting continuous sequence of pulses with clock-like periodicity,
which have been measured up to a precision of $10^{-10}$ second. Pulsar period 
varies from millisecond to few seconds and increases 
very slowly due to the loss of rotational kinetic energy of the star and 
never decreases except the occasional glitches.
  A striking observation \cite{1,2} in the study of pulsars is the  
 sudden increase in the rotational rate, 
known as 'glitches', followed by a recovery back towards the
pre-glitch period. Glitches are usually seen in young pulsars ( characteristic
age $T\sim 10^3 $ to $10^4$ years). The fractional change in the rotational 
frequency $\frac{ \Delta \Omega}{\Omega}$ during a glitch is the 
 of the order of $( \frac{\Delta\Omega}{\Omega}) \sim {10^{-9}}$ 
to ${10^{-6}}$ accompanied by jumps in the spin down rate of magnitude
$(\frac{\Delta \dot{\Omega}}{\dot{\Omega}}) \sim {10^{-3}}$ to ${10^{-2}}$.
These observations have stimulated considerable interest in the dynamics of
neutron star, particularly their interior structures. Although the radius
, mass and core structure of a neutron star are somewhat uncertain
as they depend on the equation of state(EOS) of nuclear matter for densities above the
nuclear matter density $\rho _0 \sim 2\times 10^{14} $ gm/cc, all models
of neutron star assume a solid outer crust consisting primarily of iron
nuclei arranged in 
  a very rigid and strong crystalline lattice and a neutron superfluid
coexisting with the extension of the outer crust lattice forming the inner
crust. 

Glitches can be explained \cite{3}by the presence of this superfluid component
 which is loosely coupled to the nuclei present in the rigid outer
crust. As expected \cite{4}, a rotating superfluid develops an array of quantized 
vortices, parallel to the axis of rotation, around which the circulation
 is quantized. The currently accepted explanation of pulsar glitches is
provided by the vortex creep model(VCM) \cite{5}. This model assumes that  
the superfluid vortex line within the neutron star experience a pinning
force from the crustal lattice as they move radially outward from the
rotation axis, the superfluid looses it's angular velocity and slows down.
The pinning force in this model is required to be finite and  non-uniform
so that the vortices are loosely pinned and inhomogeneously distributed
in the inner crust.
Thus a small area inside the neutron star can have a high density of
pinned vortices known as `traps' surrounded by relatively vortex-free region.   
When the vortex density within a trap increases to a 
critical value, the trap starts releasing vortices.
These vortices trigger similar release from other
traps leading to an avalanche of such vortex lines.
 As each released vortex moves away from the trap, it carries
away angular momentum from the superfluid component and transfers it to 
the crustal lattice. This spins up of the crust and results in 
glitches.
The post-glitch behavior is a outcome of formation of new vortices and 
their re-pinning to the inner crust. VCM, however, has certain
limitations. The model can not explain why very young pulsars like Crab
exhibits a different type of glitch than that of Vela. Furthermore, 
some recent works \cite{7} have raised doubt about  the existence of 
nucleon superfluidity within the neutron star core. 
 In addition, it has been suggested that the 
observed glitch and post-glitch behavior of neutron star 
to be related to its interior core structure. 
Many exotic states of matter have been proposed \cite{8,9,10}
to exist in the very dense core of the neutron star including pion-condensate,
kaon condensate, hyperon condensate, quark and gluonic matter etc.
In view of these results,  we have attempted to explain pulsar glitch
behavior in the framework of VCM considering   gluon condensates
as constituents of neutron star core.  

From QCD studies, it is proposed \cite{11} that in a QGP quraks and gluon may have 
two different phase transition temperatures, $T_g\sim 400$ MeV and $T_q\sim
250$ MeV respectively. 
Gluons are expected to undergo a first order phase transition and 
condensate to glue balls. Interestingly, Ellis et al \cite{10}  have shown 
that glue balls can be formed inside a neutron star core. 
  These  glue balls can 
transform to  hadronic matter as expected from  QCD. 
This picture of formation of hadronic
matter from QGP motivates us to propose that, while the composition of the crust
of young ($\tau\le 10^4$ years) and old ($\tau>10^4$ years) pulsars
remain the same, the core of the young pulsars consists of glue balls 
contrary to the core of the old ones, which contain hadronic matter.

 Following the arguments of Ellis et al, one may expect glueballs at the core of neutron star.
 The nature and stability of such glueballs need further detailed study. However,
from a phenomenological point of view if we accept that such glueball
obey typical QCD phase diagram \cite{12} one expects the glueballs to be stable
at the central core but decays near the boundary of core and  crustal region.
Recent studies \cite{13} have revealed various decay modes of glueballs $(\xi)$ 
including $ \xi \rightarrow p\bar{p}$ and lighter hardrons $( \pi\pi, K\bar{K})$.
 However, the widths of a glueball decay into two pions or kaons
are found to be narrow \cite{14}.
If $\xi\rightarrow p\bar{p}$ is the dominant  mode of decay then the formed
protons within the neutron star are expected to undergo electron capture and form neutrons \cite{15}.
Hence, a neutron star with glueball core have the following structure : there will
be a solid outer crust; beneath the outer crust will be an inner crust
 of superfulid neutrons and at the interior core there will be  glueballs.
 At the boundary of the inner crust and the core, glueballs  neutronize. 
With time, neutron star cools and temperature of the stellar interior
decreases \cite{16}. The decrease in the interior temperature of the neutron star allows more glueballs
to neutronize even at higher densities \cite{12}. Hence, an older pulsar is 
expected to have more neutrons than a younger one.
We now show that the pinning energy of a superfluid
vortex line to a nucleus with outer crust, is less for younger pulsars
than older ones leading to frequent glitches for the former. 

 The pinning energy $E_P$ in the VCM \cite{5} is given by :

\begin{equation}
 E_P = \frac{3}{8 \pi} [ (\frac{{\Delta ^2 }n}{E_F})_{out}  -  (\frac{{\Delta ^2 }n}{E_F})_{in}]V 
\label{1} 
\end{equation}

where $\Delta$ is the gap energy, $E_F$ is the Fermi energy and $n$
is the number density of the superfluid neutrons inside and outside the nucleus 
respectively, V  denotes the nuclear volume.
The difference between the two terms in the right hand side
of Eq. \ref{1} represents the gain in energy when the crustal nucleus is
well inside the vortex line.

To compare the pinning energies $E_{P_g}$ for younger pulsars
with glue ball core and $E_{P_n}$ for old pulsars with neutron core
,  we consider the ratio:  

\begin{equation}
 \frac {E_{P_n}}{E_{P_g}} = \frac{[ (\frac{{\Delta_n ^2 }n_n}{E_{F_n}})_{out} 
 -  (\frac{{\Delta_n ^2 }n_n}{E_{F_n}})_{in}] V_n}{ [ (\frac{{\Delta _g^2 
}n_g}{E_{F_g}})_{out}  -  (\frac{{\Delta_g ^2 }n_g}{E_{F_g}})_{\mbox{in}}]V_g} 
\label{2} 
\end{equation}
with

\begin{equation}
E_{F_{n,g}} = {\frac{\hbar^2}{2m}}{[\frac{6 \pi^2 n_{n_g}}{(2s^2 +1)}]}^{2/3}  
\label{3} 
\end{equation}
where $\hbar$ is the modified Planck constant, $m $ is the mass of the neutron, $s$
 is its spin and  
subscripts $n$ and $g$ to physical quantities refer to those appropriate
of older and younger pulsars.

Assuming $\Delta_n = \Delta_g$ and $V_n$ $=$ $V_g$,  
and neglecting the superfluid neutron number density outside
the crustal nucleus, we obtain

\begin{equation}
 \frac {E_{P_n}}{E_{P_g}} = [ \frac{n_n}{n_g} ] ^{ 1/3 }  
\label{4} 
\end{equation}

Considering that stellar core contains neutrons alone, we find the number density
of neutrons in a $ 1.33 M_{\odot}$ pulsar of radius $10 Km$ will be:  
$n_n = 3.8 \times 10^{38} $ per cc.

If the stellar core contains glue balls with typical  stellar central
density $1.25\times 10^{15}$ gm/cc a simple estimate shows that the expected  
 the number density of neutrons can be given by 
$n_g = 0.65 \times 10^{35} $ per cc.
 Hence, from equation (\ref{4}):    

\begin{equation}
 \frac {E_{P_n}}{E_{P_g}} = 18.5 
\label{10} 
\end{equation}
 This shows, the pinning energy can be substantially reduced if the
pulsar core contains glue balls instead of neutrons. 

As we have already stated, the  existence of glue ball 
in the young neutron stars helps in explaining the observed
glitch phenomenon. In the frame work of vortex creep model, glueballs 
provide weak pinning and does not affect the 
nature of glitches and  post glitch behavior (as explained by VCM).
For older pulsar, glue balls evolve into hadronic matter and
pinning becomes so strong that glitches stop while for new born 
pulsars, pinning may be super weak due to insufficient number of neutrons
and may not be observed. For very young pulsars like Crab, glitches can be
observed but the strength will be very low. For pulsars like Vela, glitches
of observed strength can be expected.

{\bf Acknowledgement: }  The author is grateful to C. S. Shukre,  A. M.
Srivastava,  A. Abbas,  A. K. Ray and  B. P. Mandal for many
helpful discussions and acknowledges financial support from Raman
Researh Institute.


\begin{references}
\bibitem{1} Radhakrishnan V and  Manchester R. N, Nature {\bf 222}, 228 (1969).
\bibitem{2} Lyne A. G, Pritchard R. S. and Shemar S. L, J. Astrophys. Astr.
{\bf 16}, 179 (1995) and references therein.
\bibitem{3} Anderson P. W and Itoh N, Nature, {\bf 256} 25 (1975);  
Ruderman M. A,
 Astrophys J {\bf 203}, 213, (1976).
\bibitem{4} Landau L. D. and Lifshifz E. M., Statistical Physics Vol I (
Butterworth -Heinemann ) (1996).
\bibitem{5} Alpar M. A, Chau H. F, Cheng K. S and Pine D, ApJ {\bf 459}, 706 
(1996) and references therein.
 \bibitem{7} Takatsuka T. and Tamagaki R, Prog. Theo. Phys {\bf 97}, 345 (1997).
\bibitem{8} Bayn G. ` Structure and Evolution of Neutron star', Editied by
Pine D. et al (Addison- Wesley), Page 188 (1992).
\bibitem{9} Heiselberg H, Perthick C. J. and Staubo E. F, Phy. Rev. Lett. {\bf 70}
, 1355 (1993).
\bibitem{10} Ellis J, Kapusta J. I and Olive K. A, Nucl. Phy. {\bf B348}, 345 (1991).
\bibitem{11} Shuryak E, Phy. Rev. Lett. {\bf 68}, 3270 (1992).
 \bibitem{12} Alam J, Raha S. and Sinha B, Phys. Rep. {\bf 273} , 243 (1996).
\bibitem{13} Bai. J , Phy. Rev. Lett. {\bf 76} 3502(1996).
\bibitem{14} Cao. J, Huang T. and Wu H. hep-th/9710311 (1997).
 \bibitem{15} Shapiro S. L. and Tenkolshy S. A , `` Black hohes, white
drarfs and neutron star" ( Wiley Interscience )  (1996). 
\bibitem{16} Pethick C. J, Rev. Mod. Phy. {\bf 64}, 1163 (1992).

\end {references}
\end{document}